\documentstyle[12pt]{article}
\newcommand{\be}{\begin{equation}}
\newcommand{\eq}{\end{equation}}
\newcommand{\ov}{\overline}

\newcommand{\Zzahl}{ {\bf Z} }

\newcommand{\Rzahl}{ {\bf R} }

\newcommand{\wt}{\widetilde}
\newcommand{\wh}{\widehat}
\begin{document}
\bibliographystyle{plain}
\renewcommand{\thefootnote}{\fnsymbol{footnote}}
\title{
{\vspace{-3cm} \normalsize \hfill MS-TPI-92-20
                                            }\\[25mm]
Critical Wilson Lines in Toroidal Compactifications of Heterotic
Strings\footnote{This work is part of a Ph.D. thesis in preparation
at the science faculty of the university of M\"unster.}}
\renewcommand{\thefootnote}{\arabic{footnote}}
\setcounter{footnote}{0}
\author{Thomas Mohaupt \\
        Institut f\"ur Theoretische Physik I,
        Universit\"at M\"unster\\
        Wilhelm-Klemm-Str.~9, D-4400 M\"unster, Germany}
\date{July 16, 1992}
\maketitle
\begin{abstract}
Critical values of Wilson lines and general background fields for
toroidal compactifications of heterotic string theories are constructed
systematically using Dynkin diagrams.

\end{abstract}
\newpage

\section{Introduction}
Toroidal compactifications \cite{GSW}, \cite{Nar}, \cite{NSW}
offer the simplest way to construct
heterotic string theories in four space--time dimensions. Although
they do not have a realistic spectrum they can be used as a
starting point for the construction of other theories like
toroidal orbifolds \cite{DHVW1}, \cite{DHVW2}, \cite{NSV}
or covariant lattice models \cite{LSW}, and some typical
features of string compactifications can be studied easily
because the inner or compactified degrees of freedom of the string
can still be represented by a free conformal field theory. String
compactifications and more generally non--minimal conformal field
theories depend on continuous parameters, called moduli, which have
been the object of many studies (see for example:\cite{NSW},
\cite{FerThe},\cite{DVV},\cite{DGH}). One is interested in the global
geometry of moduli spaces and especially in its discrete symmetries
under duality (or target space modular) transformations which in
the string interpretation are probably connected to the existence
of a minimal length scale \cite{Sha}, \cite{GRV}. Another point
of interest is the behaviour of the
spectrum and symmetry group under continuous deformations and the
existence of critical values with extended symmetry.
This has been studied systematically for those moduli of toroidal
compactifications which already appear in the bosonic case \cite{Gin},
\cite{NSW},
namely the metric $G_{ij}$ and the so called axionic background field
$B_{ij}$.
For the background gauge field (the so called Wilson lines) only
particular examples \cite{Gin} of critical values have been
mentioned so far.
The purpose of this paper is to study systematically Wilson lines
and general background fields in toroidal compactifications. This
is a first step to a better understanding of continuous
background fields, especially Wilson lines, in more general situations.
After reviewing the relevant facts about toroidal compactifications
and Lie algebras I show in detail how the Lie algebra that is present
in a toroidal compactification without Wilson lines and $B_{ij}$--field
can be broken to any maximal regular subalgebra by a suitable
choice of one Wilson line. This is done by implementing the
extended Dynkin diagram formalism, which makes the procedure
very transparent.
The results are then generalized in two steps to general
regular subalgebras of maximal and nonmaximal rank. If the
rank of the semisimple part of the Lie algebra is reduced, the
method  characterizes critical hypersurfaces instead of critical
points. Then I investigate how
the rank of the semisimple part of the gauge Lie algebra can
be enlarged beyond 16. The generalization to general background
fields does not only combine the old results about the $B_{ij}$--field
(which are reproduced) with the new results about Wilson lines
but also shows how the effects of the different background fields
can ``interfere". Finally the construction is translated into
the language of shift vectors for rational background fields.
In the last section I propose further applications
to more general compactifications.

\section{Toroidal Compactifications with Background Fields}
The simplest way to get a four-dimensional (or more generally
a D-dimensional, $D<10$) theory from the ten-dimensional
heterotic string is the compactification of six (or $d$)
space dimensions on a torus \cite{GSW}:
\be     \Rzahl^{10} \longrightarrow \Rzahl^{D} \times T^{d},
\eq
where
\be     T^{d} = \frac{\Rzahl^{d}}{2 \pi \Lambda},\;\;\;
                d = 10 - D,
\eq
and $\Lambda$ is a $d$-dimensional lattice with basis $\{{\bf e}_{i}\}$.
Since this can also be interpreted as a compactification on the
torus $\Rzahl^{d}/\Zzahl^{d}$ in the presence of a constant
background gravitational field, it is natural to admit nonzero
constant vacuum expectation values for those fields, which are
massless
space-time scalars from the D-dimensional point of view. This leads
to the Narain model \cite{Nar},\cite{NSW}, a continuous family
of string vacua parametrized
by the vacuum expectation values of the symmetric and antisymmetric
tensor field and of the gauge fields in the internal dimensions. A
state of the compactified theory is not only labeled by its
continuous space-time momentum $p^{\mu}$, $\mu = 1,\ldots,D$,
its oscillation state, and
its gauge quantum numbers $p^{I}, I=1,\ldots,16$, but also by its
winding vector
\be {\bf w} = w^{i}{\bf e}_{i} \in \Lambda,
\eq
describing how the string
wraps around the internal dimensions, and by its discrete internal
momentum
\be {\bf p} =  p_{i}{\bf e}^{*i} \in \Lambda^{*}.
\eq
$\{ {\bf e}^{*i} \}$ is the standard basis of the dual lattice
 $\Lambda^{*}$ of $\Lambda$. The metrics of the two lattices are
\be   G^{ij} =  {\bf e}^{*i}\cdot {\bf e}^{*j} \mbox{ and }
      G_{ij} =  {\bf e}_{i}\cdot {\bf e}_{j}, \mbox{   where  }
      G^{ij}G_{jk} = {\bf e}^{*i}\cdot {\bf e}_{k}  = \delta^{i}_{k}
\eq
The vectors  which consist of the gauge
quantum numbers $p^{I}$ of the ten dimensional theory
are constrained to form
a sixteen dimensional even selfdual euclidean lattice $\Gamma_{16}$
\cite{GSW}.
There are
(up to isometries) only to two such lattices: the root lattice
of the Lie algebra $E_{8} \oplus E_{8}$ and the sublattice
$D_{16}^{(0)(s)}$ of the weight lattice of the Lie algebra
$D_{16}$ that is generated by the roots and the spinor weights \cite{Cor}.
The corresponding gauge groups (of the ten dimensional theory)
are $E_{8} \otimes E_{8}$ and $Spin(32) / \Zzahl_{2}$.
All the internal quantum numbers  of the D--dimensional theory
can be combined into the left-- and right--moving momenta, which
form the momentum lattice $\Gamma \equiv \Gamma_{16+d;d}$:
\be \left( {\bf p}_{L} ; {\bf p}_{R} \right) =
     \left( p_{L}^{I} {\bf e}_{I} + p_{L}^{i} {\bf e}_{i} ;
       p_{R}^{i} {\bf e}_{i}  \right) \in \Gamma_{16+d;d},
\eq
where $\{ {\bf e}_{I} \}$ is a basis of $\Gamma_{16}$, and \cite{Gin}
\be     p_{L}^{I} = p^{I} + A^{I}_{i}w^{i},
\eq
\be        p_{L}^{i} =  \frac{1}{2}p^{i} - B_{\;k}^{i}w^{k}
        - \frac{1}{2} p^{J}A^{Ji} - \frac{1}{4}
        A^{Ki}A^{K}_{j}w^{j} + w^{i}
\eq
\be        p_{R}^{i} =  \frac{1}{2}p^{i} - B_{\;k}^{i}w^{k}
        - \frac{1}{2} p^{J}A^{Ji} - \frac{1}{4}
        A^{Ki}A^{K}_{j}w^{j} - w^{i}
\eq
In these formulas $B_{ij}$ are the components of the antisymmetric
tensor field and $A^{I}_{i}$ are the components of the background
gauge field: There is one nontrivial gauge field configuration
(called Wilson line)
${\bf A}_{i}= A^{I}_{i} {\bf e}_{I}$ for each generator of the
fundamental group of the torus
and all those fields must take values in a Cartan subalgebra
(i.e. they all commute), that is in $\Rzahl^{16}$ \cite{GSW},\cite{NSW}.
The dependence  of $\Gamma$
on the lattice $\Lambda$ is implicit in the component notation
in that the $i,j,\ldots$ indices are raised and lowered by the
metric $G_{ij}={\bf e}_{i}\cdot {\bf e}_{j}$ of the lattice $\Lambda$.
The momentum lattice is, by construction, selfdual and even with
respect to the signature $(+)^{16 +d}(-)^{d}$; this is, as
shown by Narain \cite{Nar}, equivalent to modular invariance of the
theory. $G_{ij}$, $B_{ij}$ and $A^{I}_{i}$ parametrize all $SO(16+d;d)$--
boosts of $\Gamma$ modulo proper rotations and, because there
exists up to isometries only one self dual even lattice for
each signature $8n$, $n= 1,2,\ldots$, one gets all models by
varying these parameters\footnote{Rotations are symmetries of the
space of theories, and there are further, discrete identifications
of theories due to duality \cite{NSW}.} \cite{NSW}.

The vectors of the momentum lattice contain the gauge quantum
numbers of the states. To each vector of the form
\be   ({\bf p}_{L};0) \in \Gamma_{16+d;d},\;\;{\bf p}_{L}^{2} = 2,
\eq
called a root\footnote{This useful terminology was introduced
in the context of the covariant lattice construction \cite{LSW}.}
of $\Gamma$, corresponds a charged, massless
spin 1 state, that means a charged gauge boson. These states combine
with $16 + d$ uncharged gauge bosons, which are present for
all choices of $\Gamma$, to form the adjoint representation of
the gauge group of the model under consideration, which is
of the form
\be   G = G^{(l)} \otimes U(1)^{16 +d -l},
\eq
where $G^{(l)}$ is a simply laced, semisimple Lie group of
rank $l$ and $l, 0\leq l \leq 16 + d$ is the number of linear independent
roots. If one sets $B_{ij} =0$ and $A^{I}_{i} =0$ (``compactification
without background fields") the gauge groups are, according to the
choice of the ten dimensional theory
\be  G = E_{8} \otimes E_{8} \otimes U(1)^{d} \mbox{   or   }
     G = Spin(32) / \Zzahl_{2} \otimes U(1)^{d}.
\eq
Before we can study the effects of nonvanishing background fields
on the gauge group systematically, we must recall some elements
of Dynkins theory of subalgebras of semisimple Lie algebras.

\section{Mathematical preliminaries}
A semisimple Lie algebra\footnote{The material presented here can
be found in the book by Cahn \cite{Cah}, as far as semisimple
Lie algebras are concerned. A good reference for general Lie algebras
is \cite{Hum}. A useful collection of data of simple Lie algebras
is given in appendix F of the book \cite{Cor}.}
 ${\bf g}$ of rank $l$ can be decomposed
into a Cartan subalgebra ${\bf h}$ and the one dimensional root spaces
${\bf g}_{\alpha}$, which are the eigenspaces with respect to the
adjoint action of ${\bf h}$ on ${\bf g}$:
\be    {\bf g} = {\bf h} \oplus \bigoplus_{\alpha \in \Sigma}
        {\bf g}_{\alpha}
\eq
The root spaces are labeled by a finite set $\Sigma$ of l--dimensional
vectors, whose components are the eigenvalues of the Cartan generators
in the corresponding root space. From $\Sigma$ one can choose
a set $\Pi = \{ \alpha_{i} : i=1,\ldots,l \}$ of simple roots,
from which all roots and the Lie algebra itself can be
reconstructed\footnote{The same
amount of information is encoded also in the Cartan matrix and
in the Dynkin diagram.}.

A subalgebra ${\bf g}'$ of ${\bf g}$ is called {\bf regular},
if it respects
this decomposition in the sense that its decomposition fulfills
${\bf h}' \subset {\bf h}$ and $\Sigma' \subset \Sigma$.
Non regular subalgebras
will be ignored in the following, because the method that I will
describe in the next section works by reducing the set of roots
and so only gives regular subalgebras. As the rank of the gauge
group is fixed (to be $16 + d$) in toroidal compactifications
we can restrict ourselves to regular subalgebras of maximal rank.
In order to construct subalgebras in a systematic way, one defines
${\bf g}'$ to be a {\bf maximal} subalgebra of ${\bf g}$, if
\be   {\bf g}' \subset \tilde{{\bf g}} \subset {\bf g}   \Longrightarrow
      \tilde{{\bf g}} = {\bf g}' \mbox{   or   }
      \tilde{{\bf g} } = {\bf g}
\eq
for all subalgebras $\tilde{\bf g}$. This notion needs further
refinement because a maximal subalgebra of a semisimple
Lie algebra is not necessarily semisimple. A ``maximal
semisimple subalgebra" is thus a subalgebra which is maximal
among the semisimple subalgebras. The only non semisimple
subalgebras that will appear in the following are of
the form semisimple $\oplus$ abelian, because other types of
subalgebras cannot be realized\footnote{These other types of subalgebras
have root systems which do not contain the negatives of some roots, so
they have an nilpotent part.
A representation of such an algebra cannot be constructed by the
lattice technique
used in toroidal compactifications, because the negative of a
lattice vector is always a lattice vector itself,
and also violates CPT-Invariance,
because some antiparticles are missing.}.

Dynkin has given a (not entirely correct, see below) rule to
construct all maximal semisimple subalgebras of a simple
Lie algebra. (This extends to semisimple Lie algebras because they are
direct sums of simple ones.)
The rule can be stated diagrammatically by saying that {\em one has
to remove, in all possible ways, one dot from the extended
Dynkin diagram of the simple Lie algebra} ${\bf g}$. This means, in
algebraic terms, that one chooses a system of simple roots
(which can be represented by a Dynkin diagram), then extends this
by the lowest root (thus extending the Dynkin diagram) and finally
deletes one root. The resulting set is the system of simple
roots of a maximal rank, maximal semisimple regular subalgebra of ${\bf g}$.
(There are five exceptional cases in which the subalgebra constructed
this way is not maximal. See (\ref{Exc})) This subalgebra
does not need to be
proper, but one gets all maximal semisimple regular subalgebras and by
iteration of the procedure all maximal rank, regular semisimple
subalgebras\footnote{It can also happen that one gets different but
conjugated subalgebras while generating all these algebras. The
uniqueness problem will be discussed later.}.

In order to get the
regular semisimple subalgebras of non maximal rank, one has to
combine this rule with a second one \cite{Cah}, \cite{LorGru}:
{\em Removing $k$ dots, $0<k<l$
 from the Dynkin diagram of a simple Lie algebra of rank $l$
gives the Dynkin diagram of a regular semisimple subalgebra of
rank $l-k$, and all regular semisimple subalgebras of this rank
can be constructed by removing, in all possible ways, $k$ dots
from all rank $l$ regular semisimple subalgebras}\footnote{Isomorphic
subalgebras need not be conjugated, if the Lie algebra has
outer automorphisms \cite{LorGru}.}.

In our application to the Narain model the total
rank of the gauge group is fixed, as explained above. So we will
only be able to reduce the rank of the semisimple part by removing roots,
while the Cartan subalgebra is fixed, thus giving regular subalgebras of
the type semisimple $\oplus$ abelian.

\section{Breaking to maximal Subalgebras by one Wilson line}
\subsection{Formulation of the symmetry breaking problem}
The discussion of symmetry breaking by Wilson lines starts
naturally with a compactification on a lattice $\Lambda$ with
``no background fields"
\footnote{Remind that the specification of the background
metric
$G_{ij}$ is equivalent to the specification of the lattice $\Lambda$.}
, that is with $B_{ij}= 0$ and
$A^{I}_{i} =0$.
The momentum lattice $\Gamma$ is then of the form
\be     \Gamma_{16 +6;6} = \Gamma_{16} \oplus \Gamma_{6;6},
\eq
where $\Gamma_{16}$ is one of the two sixteen-dimensional
euclidean even selfdual lattices mentioned above and $\Gamma_{6;6}$
is a twelve-dimensional lorentzian even selfdual lattice with zero
signature, which is orthogonal to $\Gamma_{16}$. The momentum
lattice has a natural basis \cite{Gin} that is given by the
basis of $\Lambda$,
$\Lambda^{*}$ and $\Gamma_{16}$:
\[     {\bf k}^{i} = \left( 0, \frac{1}{2} {\bf e}^{*i};
                       \frac{1}{2} {\bf e}^{*i}  \right)
\]
\be     \ov{{\bf k}}_{i} = \left( 0, {\bf e}_{i};- {\bf e}_{i} \right)
\label{BIGOHF}
\eq
\[     {\bf l}_{A} = \left( {\bf e}_{A}, 0;0 \right)
\]
The nonvanishing (lorentzian) scalar products between these basis vectors
are:
\be     {\bf k}^{i} \ov{{\bf k}}_{j} = \delta^{i}_{j},\;\;\;\;
         {\bf l}_{A} {\bf l}_{B} = g^{(16)}_{AB}
\label{SPIG}
\eq
where $g^{(16)}_{AB}$ is the metric of $\Gamma_{16}$.
The choice of this parametrization is very useful, because
the switching on of background fields corresponds to an isometry
of the momentum lattice and thus preserves the relations (\ref{SPIG}).
What changes is the relation between the momentum lattice and
the other lattices\footnote{In canonical operator quantization
a shift of background fields induces a shift of the canonical
momentum relative to the kinetic momentum \cite{NSW}.} \cite{Gin}:
\[     {\bf k}'^{i} = \left( 0, \frac{1}{2} {\bf e}^{*i};
                      \frac{1}{2} {\bf e}^{*i} \right)
\]
\be     \ov{{\bf k}}'_{i} = \left( {\bf A}_{i},
                      - B_{ji}{\bf e}^{*j} - \frac{1}{4}
                      ({\bf A}_{j} \cdot {\bf A}_{i} ) {\bf e}^{*j}
                      + {\bf e}_{i};
                      - B_{ji}{\bf e}^{*j} - \frac{1}{4}
                      ({\bf A}_{j} \cdot {\bf A}_{i} ) {\bf e}^{*j}
                      - {\bf e}_{i} \right)
\label{newbasis}
\eq
\[     {\bf l}_{A} = \left( {\bf e}_{A}, - \frac{1}{2}
                  ( {\bf e}_{A} \cdot {\bf A}_{i}) {\bf e}^{*i};
        -\frac{1}{2} ( {\bf e}_{A} \cdot {\bf A}_{i}) {\bf e}^{*i}
                  \right)
\]
The vectors ${\bf A}_{i} = $$A_{i}^{I}{\bf e}_{I}$ are called Wilson lines.
In order to learn about the patterns of symmetry breaking it is
useful to study the effect of one single Wilson line by setting
$B_{ij} = 0$, ${\bf A}_{i} = 0, i \neq 1$,and taking the one direction
of $\Lambda$, to which the Wilson line is assigned, to be orthogonal
to the others. The momentum lattice
is then in general of the form $\Gamma_{17;1} \oplus \Gamma_{5;5}$.
We can now analyze how the gauge group depends on the sixteen
parameters $A^{I}_{1}$.
If also $A^{I}_{1}=0$, the roots of the momentum lattice are given
by the basis vectors $l_{A}$ $=({\bf e}_{A},0;0)$
$, A=1,\ldots,16$: When $\Gamma_{16}$
is the $E_{8} \oplus E_{8}$ root lattice, the basis
vectors ${\bf e}_{A}$ are automatically a set of simple roots
$\Pi = \{ \alpha_{A} = {\bf e}_{A} \}$
of the corresponding Lie algebra and  their integer linear combinations
of norm squared 2 are the roots. When $\Gamma_{16} = D_{16}^{(o)(s)}$
the basis consists of all but one of the simple roots of $D_{16}$
plus one spinor weight vector, and the missing simple root is
given by an integer linear combination \cite{Cor}:
\be  {\bf e}_{1} = \frac{1}{2} \sum_{A=1}^{8} \alpha_{2A -1}, \;\;\;
     {\bf e}_{A} = \alpha_{A},\; A=2,\ldots,16
\label{spinorweight}
\eq
\be  \alpha_{1} =  2 {\bf e}_{1} - \sum_{A=1}^{7} {\bf e}_{2A + 1}
\eq
 As we are at the
moment only interested in the gauge group, but not in the complete
spectrum, we will only consider the sublattice generated by the
simple roots. So we find that in both cases there is a
set of vectors
\be     \widehat{\alpha}_{A} = (\alpha_{A},0;0)
\eq
which may be called simple roots of $\Gamma$ because they are a
generating set for the roots of $\Gamma$ (which, remember,
correspondent precisely to the charged gauge bosons). If we now
switch on our Wilson line ${\bf A}_{1}$ these roots are mapped to
\be   \widehat{\alpha}'_{A} = \left( \alpha_{A},
         -\frac{1}{2}(\alpha_{A}\cdot {\bf A}_{1}){\bf e}^{*1};
         -\frac{1}{2}(\alpha_{A}\cdot {\bf A}_{1}){\bf e}^{*1} \right)
       = \widehat{\alpha}_{A} -  (\alpha_{A}\cdot {\bf A}_{1}) {\bf k}^{1}
\eq
Since ${\bf k}^{1} \equiv {\bf k}'^{1}$ and the momentum lattice
 is selfdual  for all values of the background fields,
we deduce that $\widehat{\alpha}_{A}$ {\em is in the new momentum
lattice $\Gamma'$ if and only if $\alpha_{A}\cdot {\bf A}_{1}$
is an integer}\footnote{Note that all vectors in $\Gamma'$ must
have mutual integer scalar products and that $\widehat{\alpha}_{A}$
and  $\widehat{\alpha}_{A}'$ cannot be in the same selfdual lattice,
if this condition is not fulfilled. }.
Combining this observation with Dynkins theorem on maximal regular
semisimple maximal rank subalgebras we can formulate the conditions
the Wilson line ${\bf A}_{1}$ must obey in order to break the
Lie algebra ${\bf g}$ of the gauge group to such an subalgebra:
Obviously we must project out one of the simple roots, while
keeping the others {\em and the lowest root $-\alpha_{h}$}.
We can now state the symmetry breaking problem:

{\em If the Wilson line ${\bf A}_{1}$ shall break the gauge Lie
algebra to the maximal subalgebra that is defined by removing the
I-th dot from the extended Dynkin diagram, then it must fulfill}:
\be     \alpha_{I} \cdot {\bf A}_{1} \not\in \Zzahl    \label{Proj}
\eq
\be     \alpha_{A} \cdot {\bf A}_{1} \in \Zzahl,\;\;\;\;(A \neq I)
\label{UA}
\eq
\be     \alpha_{h} \cdot {\bf A}_{1} \in \Zzahl       \label{MUA}
\eq

\subsection{The standard solution of the symmetry breaking problem}
The next step is
to show, that these conditions can be fulfilled for arbitrary I. This is
somewhat trivial, because the equations above mean nothing but
${\bf A}_{1}$ being a weight vector of the subalgebra but not of the
algebra, but it is instructive and useful to construct an explicit
solution.  The symmetry breaking problem (\ref{Proj}) - (\ref{MUA})
will be solved for an arbitrary simple Lie algebra of arbitrary
rank $l$. This extends trivially to the semisimple case. It is not
necessary to specify the Lie algebra, and the technique used here
will be useful in the following.

Let ${\bf g}$ be a simple simply laced Lie algebra of rank $l$ and
with simple roots $\{\alpha_{1},\ldots,\alpha_{l} \}$ and lowest
root $ -\alpha_{h} $ and let ${\bf g}'$ be the regular maximal
rank semisimple subalgebra with simple roots
\be  \{\beta_{1},\ldots,\beta_{l} \} = \{ \alpha_{1},\ldots,
        \alpha_{I-1},-\alpha_{h},\alpha_{I+1},\ldots,\alpha_{l} \}
\eq
The equations (\ref{UA}) and (\ref{MUA}) imply that ${\bf A}_{1}$
is a weight vector of ${\bf g}'$ and thus is an integer linear
combination of the fundamental weight vectors $\beta^{*}_{A}$
of ${\bf g}'$:
\be     {\bf A}_{1} = \sum_{A=1}^{l}  a^{A} \beta_{A}^{*},
        \;\;\;a^{A} \in \Zzahl
\label{ExpWL}
\eq
The highest root $\alpha_{h}$ of ${\bf g}$ is a (known, see \cite{Cor})
integer linear combination of the simple roots of ${\bf g}$:
\be   \alpha_{h} = \sum_{A=1}^{l} k^{A} \alpha_{A},\;\;\;
        k^{A} \in \Zzahl, k^{A} \geq 0
\eq
These equations can be used to express the simple roots of ${\bf g}$
as rational linear combinations of the simple roots of ${\bf g}'$:
\be     \alpha_{I} = \frac{1}{k^{I}} \left( \alpha_{h} -
                     \sum_{A \neq I} k^{A} \alpha_{A} \right)
                =    \frac{1}{k^{I}} \left( \beta_{I} -
                     \sum_{A \neq I} k^{A} \beta_{A} \right)
\label{AlphaI}
\eq
\be     \alpha_{A} = \beta_{A}, \mbox{   for   }A \neq I
\eq
If $k^{I}=1$, $\alpha_{I}$ is a root of ${\bf g}'$ and the subalgebra
is not a proper one: ${\bf g}'$ $=$ ${\bf g}$. This corresponds
to those cases in Dynkins method where by removing a dot from
the extended Dynkin diagram one gets the original Dynkin diagram.
If $k^{I}>1$ the equations (\ref{UA}) and (\ref{MUA}) are identically
satisfied and it remains to check equation (\ref{Proj}). Since by
definition $\beta_{A}^{*} \cdot \beta_{B}$ $=\delta_{AB}$ we get
by substituting (\ref{ExpWL}) into (\ref{Proj}):
\be     \alpha_{I} \cdot {\bf A}_{1} = - \frac{1}{k^{I}} a^{I}
                - \sum_{A \neq I} \frac{ k^{A}  a^{A} }{ k^{I} }
\eq
Choosing simply
$ {\bf A}_{1} = \beta_{I}^{*}$, i.e. $a^{A}=\delta_{AI}$
we get
\be     \alpha_{I} \cdot {\bf A}_{I} = -\frac{1}{k^{I}} \not\in \Zzahl
\eq
because $k^{I} > 1$.
This solution is the most natural one because ${\bf A}_{1}$ is chosen
to be the only fundamental weight vector of the subalgebra that is
not a weight of {\bf g}. Therefore
\be     {\bf A}_{1} = \beta_{I}^{*}
\label{StWL}
\eq
will be called the {\em standard solution} of the symmetry breaking
problem.
The considerations of this section have shown that Dynkins formalism
can be
implemented in the Narain model in a simple and natural way.

\subsection{All solutions of the symmetry breaking problem}

The next question to be investigated is how many other solutions of
the equations (\ref{Proj}) -- (\ref{MUA}) exist. As these equations
are equivalent to the statement, that ${\bf A}_{1}$ is a weight vector
of ${\bf g}'$ but not of ${\bf g}$, the general solution can be
parametrized in the following way:
\be     {\bf A}_{1} = \beta_{I}^{*} + {\bf v}, \mbox{   where   }
        {\bf v} \in \Gamma_{W}({\bf g}')
\eq
Let now $\Gamma'$ be the lattice corresponding to the standard
choice ${\bf v} = 0$ of the Wilson line, whereas $\Gamma''$ is
some other solution of the symmetry breaking problem with
${\bf v} \not= 0$. These lattices can be compared most easily
by looking at there basis vectors. In the $E_{8} \oplus E_{8}$
case basis for $\Gamma'$ and $\Gamma''$ are given by
\be    \ov{\bf k}_{1}'= \left( \beta_{I}^{*} ,
          -\frac{1}{2}( \beta_{I}^{*} )^{2}
          \frac{1}{2}{\bf e}^{*1} + {\bf e}_{1} ;
          -\frac{1}{2} (\beta_{I}^{*} )^{2}
          \frac{1}{2}{\bf e}^{*1} - {\bf e}_{1}
                       \right)
\eq
\be     \ov{\bf k}_{1}''= \left( \beta_{I}^{*} + {\bf v},
           -\frac{1}{2}( (\beta_{I}^{*} + {\bf v} )^{2} )
           \frac{1}{2}{\bf e}^{*1} + {\bf e}_{1} ;
           -\frac{1}{2}( (\beta_{I}^{*} + {\bf v})^{2} )
           \frac{1}{2}{\bf e}^{*1} - {\bf e}_{1}
                        \right)
\eq
\be    \widehat{\alpha}_{I}' = \left( \alpha_{I}, -(\alpha_{I} \cdot
        \beta_{I}^{*} ) \frac{1}{2} {\bf e}^{*1};
       -(\alpha_{I} \cdot
       \beta_{I}^{*} ) \frac{1}{2} {\bf e}^{*1}
        \right)
\eq
\be    \widehat{\alpha}_{I}'' = \left( \alpha_{I}, -(\alpha_{I} \cdot
        \beta_{I}^{*} +\alpha_{I} \cdot {\bf v}) \frac{1}{2} {\bf e}^{*1};
       -(\alpha_{I} \cdot
       \beta_{I}^{*} +\alpha_{I} \cdot {\bf v}) \frac{1}{2} {\bf e}^{*1}
        \right)
\eq
\be     \widehat{\alpha}_{A} =\widehat{\alpha}_{A}' =
        \widehat{\alpha}_{A}''=
        (\alpha_{A},0;0),\mbox{   for   }A \neq I
\eq
\be     {\bf k''}^{i} = {\bf k'}^{i} = {\bf k}^{i}
        \mbox{   for all }i,\;\;\;
        \ov{\bf k}_{i}'' = \ov{\bf k}_{i}' = \ov{\bf k}_{i}
        \mbox{   for }i \neq 1.
\eq
(The unprimed vectors are the standard basis (\ref{BIGOHF})
of the lattice without background
fields, $\Gamma$.) We also need the images
of the highest root $\wh{\alpha}_{h}$ of $\Gamma$ in $\Gamma'$ and
$\Gamma''$:
\be     \widehat{\alpha}_{h}'=
         \wh{\alpha}_{h}+(-\alpha_{h} \cdot \beta_{I}^{*})
         {\bf k}^{1} = \wh{\alpha}_{h} + {\bf k}^{1}
\eq
\be     \widehat{\alpha}_{h}''=
         \wh{\alpha}_{h}+ (1 -\alpha_{h} \cdot {\bf v})
         {\bf k}^{1}, \mbox{   where   }
         -\alpha_{h} \cdot {\bf v} \in \Zzahl
\eq

In the $D_{16}$ case there is one basis vector $l_{1}''$ (see
(\ref{spinorweight}))    in the 16--dimensional
sector that is not a root, but constructed from a spinor weight. But
I will analyze the $E_{8} \oplus E_{8}$ case, where
the basis of  the lattices $\Gamma'$ and $\Gamma''$
only differ by  two vectors
$\ov{\bf k}_{1}''$ and $\widehat{\alpha}_{I}''$. The arguments given
below can be extended to the more complicated case by treating $l_{1}''$
the same way as $\widehat{\alpha}_{I}''$  will be dealt with below.

In order to make
the analysis as simple as possible, let us assume for the moment
that ${\bf v}$ is a weight vector of ${\bf g}$, which implies
$\alpha_{I} \cdot {\bf v} \in \Zzahl$. Then $\wh{\alpha}_{I}''$ and
$\wh{\alpha}_{I}'$ are in $\Gamma' \cap \Gamma''$ and we must only care
about $\ov{\bf k}_{1}''$.
If  ${\bf v}$ is even in the root lattice
of ${\bf g'}$  we can expand it
\be     \wh{\bf v} \equiv  ({\bf v}, {\bf 0}; {\bf 0} ) =
        \sum_{A=1}^{l} v^{A} (\beta_{A}, {\bf 0}; {\bf 0} ) \equiv
        \sum_{A=1}^{l} v^{A} \wh{\beta}_{A}, \mbox{   where }
        v^{A} \in \Zzahl,
\eq
implying $\wh{\bf v}$ $\in \Gamma'' $. As ${\bf v}$ is in the root
lattice and $\beta_{I}^{*}$ is a weight vector of ${\bf g'}$
it follows ${\bf v}^{2} \in 2\Zzahl$, ${\bf v} \cdot \beta_{I}^{*}$
$\in \Zzahl$ and
\be     \ov{\bf k}_{1}'' - \ov{\bf k}_{1}' = \wh{\bf v} -
                (\beta^{*}_{I} \cdot {\bf v} +
                 \frac{1}{2} {\bf v}^{2} ) {\bf k}^{1}
\eq
This difference being in $\Gamma''$
implies $\ov{\bf k}_{1}'$ $\in \Gamma''$. As all basis vectors
of $\Gamma''$ are in $\Gamma'$ and vice versa both lattices are
equal. If we want to construct a physically different solution of
the symmetry breaking problem, we must choose ${\bf v}$ $\not\in$
$\Gamma_{R}({\bf g'})$ and solutions differing by elements of
$\Gamma_{R}({\bf g'})$ are identical. As ${\bf v}$ must be in
the weight lattice of ${\bf g'}$ we can conclude:
{\em If ${\bf v}$ is a weight vector of {\bf g}, then the
inequivalent solutions of the symmetry breaking problem are
parametrized by
the elements of the finite factor group}
\be   \Gamma_{W}({\bf g'})/ \Gamma_{R}({\bf g'})
\eq
This group is well known in the theory of Lie groups and  Lie algebras
because it is (isomorphic to) the group of conjugacy classes of
representations of ${\bf g'}$ and the center of the universal
linear group $\exp({\bf g'})$ of ${\bf g'}$ \cite{GO},\cite{Cor},
\cite{LSW}.

Let us now have a look on $\wh{\alpha}_{I}''$ and relax the condition
on $\alpha_{I} \cdot {\bf v}$. Since
\be     \wh{\alpha}_{I}''  = \wh{\alpha} - (\alpha_{I} \cdot
        \beta^{*}_{I} + \alpha_{I} \cdot {\bf v}) {\bf k}^{1}
        = \wh{\alpha}_{I}'- (\alpha_{I} \cdot {\bf v}) {\bf k}^{1},
\eq
$\alpha_{I} \cdot {\bf v}$ $\not\in \Zzahl$, which is equivalent
to ${\bf v}$$ \not\in $$\Gamma_{W}({\bf g})$, implies $\wh{\alpha}_{I}''$
$\not\in$ $\Gamma'$
and $\Gamma'' \not = \Gamma'$. But as always ${\bf v} \in $
$\Gamma_{W}({\bf g'})$ and ((\ref{AlphaI}), no summation of I)
\be   k^{I}\alpha_{I} = -\beta_{I} - \sum_{A \neq I} k^{A}\beta_{A}
      \in \Gamma_{R}({\bf g'}),
\eq
we have
\be     \alpha_{I} \cdot {\bf v} = \frac{m}{k^{I}} \mbox{ mod }\Zzahl,
        \;\;\;\exists m \in \{1,\ldots,k^{I} \}
\label{m}
\eq
As both $\alpha_{I}$ and ${\bf v}$ are weights of ${\bf g}'$
their scalar product modulo integers is characterized by their
conjugacy classes \cite{LSW}.
Combining this with the previous result on the dependence of $\Gamma''$
on  $\ov{\bf k}''_{1}$ we  conclude:
{\em The inequivalent solutions of the symmetry breaking problem are
parametrized by
the elements of the finite factor group}
\be   \Gamma_{W}({\bf g'})/ \Gamma_{R}({\bf g'})
\eq

A further interesting observation, which will be useful later, is that
the case $m=1$ in equation (\ref{m}) is special. We then have:
\be     \alpha_{I} \cdot {\bf v} = \frac{1}{k^{I}} \Rightarrow
        \alpha_{I} \cdot \beta_{I}^{*} + \alpha_{I} \cdot {\bf v}=0
        \Rightarrow \widehat{\alpha}_{I} \in \Gamma''
\eq
That means that all roots of ${\bf g}$ are in $\Gamma''$ and the
algebra of symmetries realized is in fact ${\bf g}$. The condition
(\ref{Proj}), which makes sure that $\alpha_{I}$ is projected out,
is not fulfilled, because ${\bf A}_{1}$$ = $$\beta_{I}^{*} + {\bf v}$
$\in \Gamma_{W}({\bf g})$. If ${\bf A}_{1}$$\in\Gamma_{R}({\bf g})$
the basis vectors of $\Gamma''$ are all in $\Gamma$ implying
$\Gamma''$ $=\Gamma$. If not, the lattices have the same roots but
are different, that means they describe theories with identical
gauge group but different spectra. A necessary condition to
break ${\bf g}$ to ${\bf g'}$ is
\be
{\bf A}_{1} \in \Gamma_{W}({\bf g'}) - \Gamma_{W}({\bf g})
\eq
As ${\bf g'}$ is a maximal subalgebra and this choice defines
a proper subalgebra of ${\bf g}$ that contains ${\bf g'}$, it is
also sufficient.

We have also to remember that there are
cases where Dynkins rule does not give a maximal subalgebras.
But there are only the five following exceptions \cite{GolRot},
\cite{Cah} where the
algebra on the left is the would--be--maximal subalgebra\footnote{
Note that there is an apparent misprint in the book \cite{Cah}.}:
\be     \begin{array}{ccccc}
         A_{3} \oplus A_{3} \oplus A_{1} &\subset&D_{6} \oplus A_{1}
        &\subset& E_{7} \\
         A_{3} \oplus D_{5} &\subset&D_{8}&\subset& E_{8} \\
        A_{1} \oplus A_{2} \oplus A_{5} &\subset&A_{2} \oplus E_{6}
        &\subset& E_{8} \\
        A_{1} \oplus A_{7}  &\subset&A_{1} \oplus E_{7}
        &\subset& E_{8} \\
        A_{3} \oplus A_{1}  &\subset& B_{4}
        &\subset& F_{4} \\
        \end{array}
\label{Exc}
\eq
The fifth case is concerned with non simply laced Lie algebras and
cannot appear in our context.  The other cases are relevant to the
breaking of $E_{8}$. They will be discussed in the next section where
we will analyze how to break the gauge algebra to a general
semisimple maximal rank subalgebra.

\section{Maximal rank semisimple subalgebras}
Every maximal rank semisimple subalgebra of ${\bf g}$ will appear
in a chain of maximal semisimple maximal rank subalgebras:
\be  {\bf g} \supset {\bf g}^{(1)} \supset {\bf g}^{(2)} \supset \cdots
\label{COSA}
\eq
As the root and weight lattices of simply laced Lie algebras
satisfy
\be
\Gamma_{R} \subset \Gamma_{W} = \Gamma_{R}^{*}
\eq
each chain (\ref{COSA}) of maximal subalgebras induces a chain of lattices:
\be  \cdots \Gamma_{R}({\bf g}^{(2)}) \subset \Gamma_{R}({\bf g}^{(1)})
        \subset \Gamma_{R}({\bf g}) \subset \Gamma_{W}({\bf g})
        \subset \Gamma_{W}({\bf g}^{(1)})
        \subset \Gamma_{W}({\bf g}^{(2)}) \cdots
\eq
Then every choice
\be {\bf A}_{1} \in \Gamma_{W}({\bf g}^{(n)})
    - \Gamma_{W}({\bf g}^{(n-1)})
\label{BtMRSA}
\eq
of the Wilson line will break ${\bf g}$ precisely to the subalgebra
${\bf g}^{(n)}$, if we assume for the moment that ${\bf g}^{(n)}$
only appears in one such chain. To prove this note first that
for all roots
$\alpha$ of ${\bf g}^{(n)}$
\be   {\bf A}_{1} \cdot \alpha  \in \Zzahl  \Longrightarrow
      \wh{\alpha} \in \Gamma = \Gamma_{17;1} \oplus \Gamma_{5;5}
\eq
because a vector is in the selfdual momentum lattice if and only
if it has integer scalar products with all basis vectors. This
implies that all roots of ${\bf g}^{(n)}$ are roots of the
momentum lattice and ${\bf g}^{(n)}$ is contained in the gauge Lie
algebra. If on the other hand $\beta $ is a root of ${\bf g}^{(n-1)}$
that is not a root of ${\bf g}^{(n)}$, assume ${\bf A}_{1}$ $\cdot$
$\beta$ $\in$ $\Zzahl$, such that $\wh{\beta}$ $\in$ $\Gamma$. By
linearity $\Gamma$ then contains the root system of a Lie algebra
that is bigger than ${\bf g}^{(n)}$ and a subalgebra of
${\bf g}^{(n-1)}$. By maximality this Lie algebra must be
${\bf g}^{(n-1)}$ itself, but this is an contradiction because
${\bf A}_{1}$ must by construction have a non integer scalar product
with at least one root of ${\bf g}^{(n-1)}$. Therefore we can conclude
\be    {\bf A}_{1}  \cdot \beta  \not\in \Zzahl \Longrightarrow
       \wh{\beta}  \not \in \Gamma
\eq
This shows how to realize an arbitrary maximal rank subalgebra,
as long as it does not belong simultaneously to two
different chains of maximal subalgebras. If some subalgebra is maximal
in more than two subalgebras of ${\bf g}$ one has to exclude all
their weight lattices in (\ref{BtMRSA}).
The cases in which Dynkins rule fails are similar in that one
must check whether the Wilson line breaks ${\bf g}$ to the
would--be--maximal subalgebra or to the maximal subalgebra.
But this is most easily understood by an example, which may also
be useful to illustrate other aspects of the formalism.

{\bf Example:} Let us take the $E_{8} \oplus E_{8}$ model as
our example and ignore the second $E_{8}$. The vectors $\wh{\alpha}_{A}$,
$A=1,\ldots,8$, where $\alpha_{A}$ are the simple roots\footnote{The
conventions for roots used here are the same as in the book by
Cornwell.}
 of $E_{8}$,
are basis vectors of the momentum lattice. Looking at the extended
Dynkin diagram, we find that the lowest root is $-\alpha_{7}^{*}$.
It can be expanded in the simple root basis using the inverse
of the Cartan matrix $C(E_{8})$ of $E_{8}$:
\be   -\alpha_{7}^{*} = - \sum_{A=1}^{8} k^{A} \alpha_{A} =
        - \sum_{A=1}^{8} C(E_{8})^{-1}_{\;\;7A} \alpha_{A}
\eq
\[      =- 2 \alpha_{1} -  4 \alpha_{2} - 6 \alpha_{3} -  5 \alpha_{4} -
        4 \alpha_{5} - 3 \alpha_{6} -  2 \alpha_{7} - 3 \alpha_{8}
\]
In order to break $E_{8}$ to the subalgebra $E_{7} \oplus A_{1}$,
we have to remove $\alpha_{7}$. To choose the Wilson line we write
down the simple roots of the subalgebra:
\be     E_{7}:\; \beta_{A} = \alpha_{A}, A=1,\ldots,6, \beta_{7} =
        \alpha_{8},\;\;A_{1}:\; \beta = -\alpha_{7}^{*}
\eq
The standard choice (\ref{StWL}) for the Wilson line is
then the dual of the new
simple root $\beta$, where ``dual" refers to the subalgebra
$E_{7} \oplus A_{1}$:
\be {\bf A}_{1} = \beta^{*} = \frac{\beta}{\beta \cdot \beta}
\eq
Working out the scalar products between the simple roots and this
Wilson line shows that in fact only
\be  \beta^{*} \cdot \alpha_{7}  =  \frac{-\alpha_{7}^{*} \cdot
        \alpha_{7} }{ \alpha_{7}^{*} \cdot  \alpha_{7}^{*} } =
        \frac{-1}{2}
\eq
is non--integer. When we now look at the extended Dynkin diagram
of $E_{7}$ we find that the lowest root is $-\beta_{1}^{*}$
and that we can break $E_{7} \oplus A_{1}$ to $A_{7} \oplus A_{1}$
by removing the seventh root of $E_{7}$. The basis vectors of the
subalgebra are:
\be     A_{7}: \; \gamma_{1} = -\beta_{1}^{*} = \sum_{A=1}^{7}
        C(E_{7})^{-1}_{\;\;1A} \beta_{A},\;\;
        \gamma_{A} = \beta_{A-1}  ,A=2,\ldots,7, \;\;\;A_{1}:\;\beta
\eq
and the standard Wilson line is therefore
\be     {\bf A}_{1} = \beta^{*} + \gamma_{1}^{*},\;\;
        \gamma_{1}^{*} = \sum_{A=1}^{7} C(A_{7})^{-1}_{\;\;1A} \gamma_{A}
\eq
This subalgebra is one of the exceptions to Dynkins rule because
it is not maximal in $E_{8}$ but can be constructed directly from the
extended Dynkin diagram of $E_{8}$ by removing the second root\footnote{
These two subalgebras are isomorphic and conjugated by an inner
automorphism \cite{GolRot}.} \cite{Cah},\cite{GolRot}.
Applying the procedure to this case we choose a basis
\be     A_{7}:\; \gamma'_{1} = \alpha_{8}, \gamma'_{A} = \alpha_{A+1},
        A=2,\ldots,6, \gamma'_{7} = -\alpha_{7}^{*},\;\;A_{1}:\;
        \beta'= \alpha_{1}
\eq
and the standard Wilson line
\be   {\bf A}_{1} = (\gamma'_{7})^{*} = \sum_{A=1}^{7}
            C(A_{7})^{-1}_{\;\;7A} \gamma'_{A}
\eq
But since $A_{7} \oplus A_{1}$ is not maximal in $E_{8}$ we do not
know whether some extra roots survive the projection and extend
the root system to the one of $E_{7} \oplus A_{1}$. These extra
roots must then come from the weight  $-(\gamma'_{4})^{*}$
 of $E_{7}$ that extends the Dynkin diagram
of $A_{7}$ to the extended Dynkin diagram of $E_{7}$\footnote{This means
that $-(\gamma'_{4})^{*}$, $\gamma_{2}'$, $\ldots$, $\gamma_{7}'$ are
a system of simple roots of $E_{7}$.}.
The vector $(\wh{\gamma}'_{4})^{*} $ is in the momentum lattice
if and only if $(\gamma'_{4})^{*}$ has integer scalar product with
$\alpha_{2}$ (in order to be compatible with $\wh{\alpha}_{2}'$) and
with the Wilson line ${\bf A}_{1}$.
As
\be   (\gamma'_{4})^{*} \cdot \alpha_{2} = \sum_{A=1}^{7}
        C(A_{7})^{-1}_{4A} (\gamma'_{A} \cdot \alpha_{2}) =
        C(A_{7})^{-1}_{42}  (-1) = \frac{8}{8} (-1) = -1 \in \Zzahl,
\eq
it depends only on the choice of the Wilson line. The standard
choice $(\gamma'_{7})^{*}$ gives
\be  {\bf A}_{1} \cdot (\gamma'_{4})^{*} = (\gamma'_{7})^{*}  \cdot
        (\gamma'_{4})^{*} = C(A_{7})^{-1}_{74} =
       \frac{4}{8}  \not \in \Zzahl
\eq
and the symmetry algebra is only $A_{7} \oplus A_{1}$.

An example
of symmetry enhancement is provided by the choice ${\bf A}_{1} =$
$2(\gamma'_{7})^{*}$. This projects out the second root of
$E_{8}$, because
\be   2(\gamma'_{7})^{*}  \cdot \alpha_{2} =
      2 C(A_{7})^{-1}_{72} (-1)
        = 2 \frac{2 }{8} = \frac{1}{2} \not \in  \Zzahl.
\eq
but is compatible with the extra root $-(\gamma'_{4})^{*}$:
\be 2 (\gamma'_{7})^{*} \cdot (\gamma'_{4})^{*}
        = 2 \frac{4}{8} = 1 \in \Zzahl
\eq
In this case the gauge Lie algebra is $E_{7} \oplus A_{1}$.

\section{One parameter families}
By now we know a discrete set of critical values of the Wilson line,
corresponding to maximal rank semisimple subalgebras of ${\bf g}$.
In order to study the behaviour between these special points in
the moduli space consider the one parameter families of Wilson
lines defined by
\be     {\bf A}_{1} = p \beta_{I}^{*},\;\;\;\;p \in \Rzahl
\eq
which interpolate between the model without background fields
and the models with the standard Wilson lines $\beta_{I}^{*}$
corresponding to
the removal of the I-th dot from the extended Dynkin diagram of
${\bf g}$.
The basis vectors of the lattice $\Gamma(p)$ corresponding to this
choice of Wilson line are
\be     \ov{\bf k}_{1} '=\left( p \beta_{I}^{*},
                -\frac{1}{2} p^{2} (\beta_{I}^{*} \cdot \beta_{I}^{*})
                \frac{1}{2} {\bf e}^{*1}  +  {\bf e}_{1};
                -\frac{1}{2} p^{2} (\beta_{I}^{*} \cdot \beta_{I}^{*})
                \frac{1}{2} {\bf e}^{*1}  - {\bf e}_{1} \right)
\eq
\be     \widehat{\alpha}_{I}' = \left( \alpha_{I},
                        \frac{p}{k^{I}} \frac{1}{2} {\bf e}^{*1};
                        \frac{p}{k^{I}} \frac{1}{2} {\bf e}^{*1}
                      \right) =
        \wh{\alpha}_{I} + \frac{p}{k^{I}} {\bf k}^{1}
\eq
\be     \widehat{\alpha}_{A}' = (\alpha_{A}, {\bf0};{\bf 0} )
                         = \wh{\alpha}_{A},\;\;A \neq I
\eq
\be     \widehat{\alpha}_{h}' = \left( \alpha_{h},
                                 p \frac{1}{2} {\bf e}^{*1}
                                ;p \frac{1}{2} {\bf e}^{*1} \right)
              = \wh{\alpha}_{h} + p {\bf k}^{1}
\eq
We can now analyze the root system of this lattice for different
values of $p$, using the methods introduced in the last sections.

For $p \not \in \Zzahl$,
neither $\wh{\alpha}_{I}$ nor $\wh{\alpha}_{h}$ are in
$\Gamma(p)$ and the gauge algebra is broken to
\be    {\bf g}(p) = {\bf g}^{(15)}_{I} \oplus u(1)
\eq
where ${\bf g}^{(15)}_{I}$ is the rank 15 semisimple Lie algebra
constructed by removing the I-th dot from the Dynkin diagram of
${\bf g}$ and $u(1)$ is the one dimensional abelian Lie algebra.
(These Wilson lines solve (\ref{Proj}) and (\ref{UA}), but not
(\ref{MUA}.)

The case $p=1$ corresponds to the standard solution (\ref{StWL})
and the gauge algebra is the rank 16 semisimple Lie algebra
${\bf g}^{(16)}_{I}$  constructed
by removing the I-th dot from the extended Dynkin diagram of ${\bf g}$:
\be   {\bf g}(1) =  {\bf g}^{(16)}_{I}
\eq

If $p = k^{I}$, both  $\wh{\alpha}_{I}$ and   $\wh{\alpha}_{h}$ are
in $\Gamma(p)$:
\be   {\bf g}(k^{I}) = {\bf g}
\eq
This also applies to all integer multiples of $k^{I}$.
The lattices $\Gamma = \Gamma(0)$ and $\Gamma(k^{I})$ are equal if and
only if $\alpha_{I}^{*}$$\in \Gamma_{R}({\bf g})$. Since $\alpha_{I}^{*}$
$=-k^{I}\beta_{I}^{*}$
$\in \Gamma_{W}({\bf g})$\footnote{By construction $\beta_{I}^{*} \cdot$
$\alpha_{I}$ $=-\frac{1}{k^{I}}$ and $\beta_{I}^{*} \cdot$ $\alpha_{J}$
$=0$, for all $J \not= I$, implying $\beta_{I}^{*}=$ $-\frac{1}{k^{I}}$
$\alpha_{I}^{*}$.},
 there must be some integer $M \in \Zzahl$
such that $M\alpha_{I}^{*} \in \Gamma_{R}({\bf g})$:
\be    {\bf g}(nk^{I}) = {\bf g},\;(\forall n \in \Zzahl),\;\;\;
        \Gamma(Mk^{I}) =  \Gamma,\;(\exists M \in \Zzahl^{+})
\eq
For all integer values of $p$, that are not multiples of $k^{I}$
the gauge Lie algebra is ${\bf g}^{(16)}_{I}$. Whether $\Gamma(p)$
equals $\Gamma(1)$ or not depends on $\beta_{I}^{*} \cdot$
$\beta_{I}^{*}$ being integer or not.

For all non integer values of $p$ the rank of the semisimple part
is reduced by one. We could now study the 15--dimensional sublattice
and add a discrete piece to the Wilson line to break ${\bf g}^{(15)}$
to its
maximal rank subalgebras, then again reduce the rank and so on. The
reduction of the rank will of cause increase the number of continuous
parameters accordingly. By removing in all possible ways simple
roots from all maximal rank semisimple subalgebras of ${\bf g}$,
we can realize all subalgebras of the type semisimple $\oplus$
abelian.

We could also study one parameter families
interpolating between ${\bf g}$ and its diverse subalgebras, investigate
multiparameter families and so on.
One example of a two parameter family will be discussed later
after generalizing to general background fields.
All these considerations give some information about the Narain
moduli space that can be made more and more precise by investing
time and effort.

\section{Enhancement of symmetry by one Wilson line}
By a ``critical" or ``multicritical"  Wilson line people usually mean
values of the Wilson line that (in the Narain model) extend the rank of
the semisimple part of the gauge algebra beyond 16. This has not yet
been analyzed systematically but there is the example of Ginsparg
\cite{Gin}
that shows how $D_{16}$ can be enlarged to $D_{17}$ (or more
generally to $D_{16 + d}$, if $\geq d$ dimensions are compactified)
by a suitable choice of Wilson line(s). I will now discuss a more
general problem but restrict the analysis for the moment to lattices
of the type $\Gamma_{17;1} \oplus \Gamma_{5;5}$ that means to the
case of one nonvanishing Wilson line.

Suppose that the Wilson line is chosen such that the lattice
contains $l \leq 16$ roots. We can then look for a vector of
the form
\be     \wh{\alpha}_{l+1} = ({\bf v}, {\bf w}; {\bf 0} )
\eq
which extends the root system. Then the $l+1$ simple root vectors
must fulfill
\be    \wh{\alpha}_{A} \cdot  \wh{\alpha}_{B} =  C^{(l+1)}_{AB},\;\;\;
        A,B = 1,\ldots,l+1,
\label{Extension}
\eq
where $C^{(l+1)}_{AB}$ is the Cartan matrix of a rank $l+1$
semisimple Lie algebra. By inspection of the basis, the most
general ansatz for $\wh{\alpha}_{l+1}$ is
\be     \wh{\alpha}_{l+1} =  \ov{\bf k}_{1}  + m {\bf k}^{1}
        , \; m \in \Zzahl
\eq
\[        = \left(  {\bf A}_{1},
        -\frac{1}{2}({\bf A}_{1} \cdot
        {\bf A}_{1}) \frac{1}{2} {\bf e}^{*1} +  {\bf e}_{1};
        -\frac{1}{2}({\bf A}_{1} \cdot
        {\bf A}_{1}) \frac{1}{2} {\bf e}^{*1} -  {\bf e}_{1}
        \right) +
        m \left(  {\bf 0}, \frac{1}{2} {\bf e}^{*1};
                \frac{1}{2} {\bf e}^{*1}   \right)
\]
Since our space--time torus is $\Rzahl^{d} / 2\pi \Lambda$,
the square root of ${\bf e}_{1} \cdot {\bf e}_{1}$ is the
``radius"\footnote{The geometric interpretation is ambiguous due to
duality transformations which  map $R$ tp $\frac{1}{2R}$, but leave
the physics of the model invariant \cite{GinRev}, \cite{GRV}.}
$R$ of the 1-direction, which we have chosen  to be orthogonal to
the others. This implies:
\be    {\bf e}_{1} \cdot {\bf e}_{1} = R^{2}, \;\;\;
       {\bf e}^{*1} \cdot {\bf e}^{*1} = \frac{1}{R^{2}},\;\;\;
        {\bf e}_{1}  = R^{2} {\bf e}^{*1}
\eq
It is very useful to rewrite $\ov{\bf k}_{1}'$:
\be     \ov{\bf k}_{1}'= \left( {\bf A}_{1},
        2{\bf e}_{1} + D \frac{1}{2} {\bf e}^{*1};
                       D \frac{1}{2} {\bf e}^{*1}  \right),\;\;\:
        D = 2 \left( -R^{2} - \frac{1}{4} ({\bf A}_{1} \cdot {\bf A}_{1})
                \right)
\eq
This shows that $\wh{\alpha}_{l+1}$ can be in $\Gamma'$ only if
\be     D = -m \in \Zzahl
\label{RootinLattice}
\eq
The extra root has the form
\be     \wh{\alpha}_{l+1} = \left( {\bf A}_{1}, 2 {\bf e}_{1}; {\bf 0}
        \right)
\eq
$\wh{\alpha}_{l+1}$ being a root vector also implies:
\be    \wh{\alpha}_{l+1}  \cdot  \wh{\alpha}_{l+1}  = {\bf A}_{1} \cdot
        {\bf A}_{1} + 4 {\bf e}_{1} \cdot {\bf e}_{1}  = 2
\label{Rootnorm}
\eq
(\ref{RootinLattice}) and (\ref{Rootnorm}) imply  two conditions:
\be     2 R^{2} + \frac{1}{2}  {\bf A}_{1}  \cdot {\bf A}_{1}
        = m \in \Zzahl
\eq
\be     4 R^{2} + {\bf A}_{1}  \cdot {\bf A}_{1}
        = 2
\eq
Combinig them fixes $m=1$ and gives a relation between the radius
of the 1--direction of the torus and the norm of the Wilson line:
\be     R^{2} = \frac{1}{2} - \frac{1}{4} {\bf A}_{1}  \cdot {\bf A}_{1}
\eq
This shows that the enhancement of symmetry takes only place for
special values of the radius.
Since $R^{2} >0$ and ${\bf A}_{1}  \cdot {\bf A}_{1}$ $\geq 0$
we have the following bounds on the radius and the norm of the
Wilson line:
\be  0 < R \leq \frac{1}{2} \sqrt{2},\;\;\; 0 \leq
        {\bf A}_{1}  \cdot {\bf A}_{1} \leq 2
\eq
The restriction on the radius--parameter is afflicted by the
ambiguity of geometrical interpretation due to duality (or target
space modular) transformations \cite{GRV}, \cite{GinRev}.
As the radius of the transformed
model is $1/2R$ we cannot distinguish between very small and very
large radii and the inequality above only excludes copies.

In order to analyze the condition (\ref{Extension}) we assume that
the extra root extends the rank $l$ algebra such that it has
a nonvanishing scalar product with only one of the l simple roots,
let us say with the I--th one:
\be     \wh{\alpha}_{l+1} \cdot \wh{\alpha}_{A} = {\bf A}_{1} \cdot
        \alpha_{A} = - \delta_{AI}  \Rightarrow {\bf A}_{1} =
        - \alpha_{I}^{*} + {\bf v}
\eq
where ${\bf v}$ is in the orthogonal complement (with respect to
$\Rzahl^{16}$) of the linear hull of $\alpha_{1},$$\ldots,$$\alpha_{l}$.
If the extra root has vanishing scalar product with all other roots,
${\bf A}_{1}$ lies completely in this complement and the gauge algebra
is extended by a $A_{1}$ algebra. For ${\bf A}_{1}$ $=0$ we have
$R=$ 1/$\sqrt{2}$. This is well known from one dimensional
compactifications \cite{GSW} and $(c=1)$--CFT \cite{GinRev}.
If ${\bf A}_{1}$ shall have a nonvanishing scalar product with more
than one other root it must be an appropriate linear combination of
fundamental weights plus an arbitrary orthogonal piece.

Finally note that we now have found examples of special values
for all the 17 continuous parameters (moduli) of $\Gamma_{17;1}$,
the Wilson line ${\bf A}_{1}$ and the radius $R$.

\section{Symmetry breaking and enhancement by more then one
Wilson line}
In this section we return to the general case by allowing
arbritray background fields.
In this case all the $6+6$ dimensions of the momentum lattice that
come from compactification (the compactification sector)
can be mixed with the 16 dimensions
coming from the gauge sector of the ten--dimensional theory.
The most obvious generalization of the previous results is to
split the role of the single Wilson line used above  among the up to
six Wilson lines. As it was possible to construct arbitrary subalgebras
by one Wilson line this cannot lead to new symmetry algebras but only
to different spectra. Instead of studying this pure refinement of
the old pattern we will investigate the new patterns of symmetry breaking
and enhancement that are now available: As a first step we will
show that extra roots can be constructed for special domains of
values of the Wilson lines. The effect on the gauge algebra
coming from the 16--dimensional sector will be investigated in a
second step.
The following parametrization of the basis vectors of the
momentum lattice makes the appearance of extra roots very transparent:
\be     {\bf k'}^{i} = \left( 0, \frac{1}{2} {\bf e}^{*i};
                \frac{1}{2} {\bf e}^{*i} \right) = {\bf k}^{i}
\eq
\be     \ov{\bf k}_{i}' = \left( {\bf A}_{i}, 2{\bf e}_{i} + D_{ij}
                \frac{1}{2} {\bf e}^{*j}; D_{ij} \frac{1}{2} {\bf e}^{*j}
                \right)
\eq
\be     l_{A}'= \left( {\bf e}_{A}, -({\bf e}_{A} \cdot {\bf A}_{k})
                               \frac{1}{2}
                                {\bf e}^{*k};
                              -({\bf e}_{A} \cdot {\bf A}_{k}) \frac{1}{2}
                                {\bf e}^{*k} \right)
\eq
where
\be     D_{ij} = 2 \left(B_{ij} -  G_{ij} -\frac{1}{4}
                ({\bf A}_{i} \cdot  {\bf A}_{j}) \right)
\label{DefD}
\eq
By inspection the only way to construct root vectors out of the
${\bf k'}^{i}$ and the $\ov{\bf k}_{i}'$ is  to take as candidates
for simple roots the linear combinations
\be     \wh{\alpha}_{16+i} \equiv \ov{\bf k}_{i}' - D_{ij}{\bf k'}^{j}
        =  \left( {\bf A}_{i}, 2 {\bf e}_{i} ; {\bf 0} \right)
\label{ExtraRoots}
\eq
and these are in the momentum lattice $\Gamma'$ if and only if
\be     D_{ij}  \in \Zzahl.
\eq
Note that this condition only fixes a combination of the Wilson
lines and the six--dimensional lattice metric $G_{ij}$ which is the
symmetric part of $D_{ij}$
and leaves us free
to vary one of these objects if we vary the other accordingly.
The extra roots form the system of simple roots of a semisimple
Lie algebra of rank $\leq 6$, if
\be     \widehat{\alpha}_{16+i} \cdot \widehat{\alpha}_{16+j}
        = {\bf A}_{i} \cdot {\bf A}_{j} + 4 G_{ij} = C_{ij}
\label{EXT6}
\eq
where $C_{ij}$ is the Cartan matrix of such an algebra. In this
equation the matrices in the middle are subject to constraints:
$G_{ij}$ must be positive definite because it is a metric and
${\bf A}_{i} \cdot {\bf A}_{j}$ must be positive semidefinite
because it is a matrix of scalar products of vectors. This implies
for example that there are no admissible solutions to equation
(\ref{EXT6}) when one of the Wilson lines  has a norm squared
$\geq 2$ because then $G_{ij}$ would have a zero or negative
eigenvalue. We have also to ensure that $D_{ij}$ is an integer
because this is necessary and sufficient for the extra roots being
vectors in $\Gamma'$. But combining (\ref{DefD}) and (\ref{EXT6})
this is equivalent to
\be     B_{ij}  =  G_{ij} +\frac{1}{4} {\bf A}_{i} \cdot {\bf A}_{j}
        = \frac{1}{4} C_{ij} \mbox{   modulo  } \frac{1}{2}
\label{ConstraintB}
\eq
This condition on $B_{ij}$ can always be fulfilled although
$B_{ij}$ is antisymmetric: The Cartan matrices $C_{ij}$
of simply laced Lie algebras have entries 2 on the diagonal and
0 and $-1$ off diagonal and they are symmetric so that they are
also antisymmetric modulo 2. If one sets all Wilson lines to zero
one recovers the old results \cite{Gin} about symmetry enhancement by
the $B_{ij}$ field. On the other hand we can set $B_{ij}$ to zero and
analyze the possibility to construct extra roots through the Wilson
lines and the lattice metric alone. But if we want to have
(\ref{EXT6}) and $D_{ij} \in \Zzahl$ the Cartan matrix cannot
have $-1$'s  as offdiagonal entries and the extra roots are restricted
to form a Lie algebra that is simply a direct sum of $A_{1}$-algebras.
Combining nonvanishing background fields of all kinds we can have
symmetry breaking by Wilson lines and symmetry enhancement by
the $B_{ij}$-field and by Wilson lines. If we construct extra
roots  (\ref{ExtraRoots}) with nonvanishing Wilson lines then they
can have integer scalar products with some roots from the
16--dimensional sector, resulting in a simple summand of the
total gauge algebra that is made of roots coming from both sectors.

Each choice of Wilson lines will break the gauge algebra
${\bf g}^{(16)}$
of the sixteen dimensional sector to a subalgebra. In fact a
root of $\Gamma$ of the form
\be   \wh{\alpha}  = \left(\alpha, {\bf 0};{\bf 0}  \right)
\eq
is in the new lattice if and only if
\be   \alpha \cdot  {\bf A}_{i} \in \Zzahl  \;\:\:(i= 1,\ldots,6)
\eq
because the momentum lattice is selfdual. Of course the Wilson lines
that produce extra roots in the six--dimensional sector need not to
fulfill
the criterion for breaking  ${\bf g}^{(16)}$ to a maximal rank subalgebra
and thus will generically reduce the rank of ${\bf g}^{(16)}$. Let us
analyze the situation by an example that is typical for all models
with 2 nonvanishing Wilson lines:

{\bf Example:} We start with ${\bf g}^{(16)} = D_{16}$ and take
two nonvanishing Wilson lines, ${\bf A}_{1}$ and ${\bf A}_{2}$.
The metric $G_{ij}$ is chosen to be blockdiagonal, such that the
two directions to which the nonvanishing Wilson lines are assigned
are orthogonal to the other four directions that can therefore be
ignored.
We can then construct the two simply laced semisimple Lie algebras
$A_{1} \oplus A_{1}$ and $A_{2}$ by setting $B_{ij}$ to its critical
value (\ref{ConstraintB}) and
solving (\ref{EXT6}) for
\be     C^{(A_{1} \oplus A_{1})}_{ij}       =
        \left( \begin{array}{cc} 2 & 0 \\ 0 & 2\\ \end{array}
        \right)
        \mbox{   or    }
        C^{(A_{2})}_{ij}                     =
        \left( \begin{array}{cc} 2 & -1 \\ -1 & 2\\ \end{array}
        \right)
\eq
Both cases can be treated simultaneously.
Next we try to use our freedom in choosing ${\bf A}_{1}$ and
${\bf A}_{2}$ to
break $D_{16}$ to the subalgebra $A_{3} \oplus A_{3} \oplus D_{10}$.
This subalgebra is constructed by removing first the third root from
the extended Dynkin diagram of $D_{16}$, resulting in the maximal
subalgebra $A_{3} \oplus D_{13}$ and then removing the third root
from the extended Dynkin diagram of the $D_{13}$. If we denote by
$\gamma_{1},\ldots,\gamma_{3},$$\gamma_{4},\ldots,\gamma_{6},$
$\gamma_{7},\ldots,\gamma_{16}$ the simple roots of the
$A_{3} \oplus A_{3} \oplus D_{10}$ subalgebra then one solution of
this symmetry breaking problem is given by
\be     {\bf A}_{1} = \gamma_{3}^{*},\;\;\;
        {\bf A}_{2} = \gamma_{6}^{*}
\eq
Substituting the scalar products
\be     {\bf A}_{1} \cdot {\bf A}_{1} = \gamma_{3}^{*} \cdot
        \gamma_{3}^{*} = \frac{3}{4}
\eq
\be     {\bf A}_{2} \cdot {\bf A}_{2} = \gamma_{6}^{*} \cdot
        \gamma_{6}^{*} = \frac{3}{4}
\eq
and
\be     {\bf A}_{1} \cdot {\bf A}_{2} = \gamma_{3}^{*} \cdot
        \gamma_{6}^{*} = 0
\eq
into equation (\ref{EXT6}) we find that we have to choose
\be     (G_{ij})_{i,j=1}^{2} = \frac{1}{16}
                   \left( \begin{array}{cc} 5 & 0 \\0 & 5 \\
                  \end{array} \right)
\eq
and
\be     (G_{ij})_{i,j=1}^{2} = \frac{1}{16}
                   \left( \begin{array}{cc} 5 & -4 \\-4 & 5 \\
                  \end{array} \right).
\eq
Both solution
are positive definite and therefore admissible. Calculating the
scalar products between the extra simple roots
\be     \widehat{\gamma}_{17} = -(\gamma_{3}^{*}, 2 {\bf e}_{1} ; 0)
\eq
\be     \widehat{\gamma}_{18} = -(\gamma_{6}^{*}, 2 {\bf e}_{2} ; 0)
\eq
and the simple roots $\wh{\gamma}_{A}$, $A=1,\ldots, 16$ of
$A_{3} \oplus A_{3} \oplus D_{10}$
\be     \widehat{\gamma}_{A} \cdot \widehat{\gamma}_{17} =
        -\gamma_{A} \cdot \gamma_{3}^{*} = -\delta_{A3}
\eq
\be     \widehat{\gamma}_{A} \cdot \widehat{\gamma}_{18} = -
        \gamma_{A} \cdot \gamma_{6}^{*} = -\delta_{A6}
\eq
we find that these roots are linked together such that the total
symmetry algebra is $A_{4} \oplus A_{4} \oplus D_{10}$ in the first
and $A_{8} \oplus D_{10}$ in the second case. This example shows
how symmetry breaking and symmetry enhancement can interfere. Note
that the semisimple part of the new Lie algebra that has a
bigger rank then the
one we started with although we did not simply extend the original
algebra by simple roots. We have also shown that one can construct ``big"
simple pieces, like $A_{4}$ and $A_{8}$ by combining roots from
the gauge sector and the compactification sector.

We can now study two parameter families of theories by setting
\be    {\bf A}_{1} = p_{1} \gamma^{*}_{3},\;\;
       {\bf A}_{2} = p_{2} \gamma^{*}_{6}
\eq
and varying the $p_{i}$. If both $p_{i}$ $\not \in \Zzahl$
the roots $\gamma_{3}$ and $\gamma_{6}$ are projected out and the
symmetry is reduced to
\be  A_{2} \oplus A_{2} \oplus D_{10} \oplus \left\{
        \begin{array}{l}  A_{1} \oplus A_{1}  \\ A_{2}  \end{array}
        \right.
\eq
respectively, that is we get critical surfaces of rank 16 models in
the Narain moduli space. If we restrict one $p_{i}$ to be an integer
we get critical lines of rank 17 models within these surfaces. These
lines intersect  each other in critical points representing
rank 18 models when
both $p_{i}$ are integers. We can try to learn about the global
geometry of this surface by analyzing the constraint that the
metric $G_{ij}$ must be positive definite. This fixes the domain
of the parameters to be
\be   0 < p_{i}^{2} < \frac{8}{3},\;\;
        i = 1,2.
\eq
in the $A_{1} \oplus A_{1}$ case and   to be
\be      0< p_{1}^{2} < \frac{8}{3}  \mbox{   and   }
        \left( 2 - \frac{3}{4} p_{1}^{2} \right)
        \left( 2 - \frac{3}{4} p_{2}^{2} \right) - 1 >0
\eq
in the $A_{2}$ case.

Although we analyzed a specific example it is clear that the pattern
found here is generic for models with more then one Wilson line.
One interesting generalization of this example is to take the
space--time torus to be a product of three two--dimensional tori, where
each of these is proportional to the maximal torus of $SU(3)$ and
six Wilson lines. This region of the Narain moduli space is the
natural starting point for the construction of $\Zzahl_{3}$--orbifolds
with continuous Wilson lines. (Of course the background fields must
be further constraint, so that the orbifold twist is an automorphism
of the momentum lattice. This and related questions are under
investigation.)

The example suggests the question whether it is possible to
combine every symmetry breaking by Wilson lines with every
system of extra roots generated by the $B_{ij}$ field.
To give a counterexample, it is sufficient to study the simplest
nontrivial case: Assume that there is  one nonvanishing Wilson
line ${\bf A}_{1}$ and that the extra root
$({\bf A}_{1}, 2{\bf e}_{1}; {\bf 0})$ has a nonvanishing
scalar product with precisely one of the other extra roots
$({\bf 0},2{\bf e}_{i};{\bf 0})$, $i=2,\ldots,l \leq 6$.
Then the only nontrivial constraint is that the matrix
\be     G_{ij} = \frac{1}{4}(C_{ij} - {\bf A}_{i}\cdot{\bf A}_{j})
        =\frac{1}{4} \left(
        \begin{array}{cccc}
                2-{\bf a}^{2}&-1 &0 &\cdots \\
                       -1      &2&* & \\
                        \vdots     &* &\ddots& \\
                             & &      &2\\
        \end{array}
        \right)
\eq
must be positive definite, where ${\bf a}^{2} = {\bf A}_{1} \cdot$
${\bf A}_{1}$ $>0$. The matrix that we get by removing the first
row and the first column is the Cartan matrix of a Lie algebra and
therefore positive definite. So we have only to look for the determinant
of $G_{ij}$ and this can be done case by case: The Lie algebra
${\bf g}^{(6)}$ formed by the extra roots must be one of the
Lie algebras
\be     A_{l} \oplus X_{6 - l}, \;\; 1 \leq l \leq 6,\;\;\;
        D_{l} \oplus X_{6 - l}, \;\; 4 \leq l \leq 6,\;\;\;
        E_{6}
\eq
where the ideal $X_{6-l}$ is orthogonal to the ideal to which the
first root belongs.
The determinant
of $G_{ij}$ can be expanded and expressed as a function of
${\bf a}^{2}$ and the determinants of the Cartan matrices of the
subalgebras of ${\bf g}^{(6)}$. The result is an upper bound on
${\bf a}^{2}$
\be     \det (G_{ij}) > 0 \Leftrightarrow
        {\bf a}^{2} < \left\{ \begin{array}{cc}
                \frac{l+1}{l},  &  {\bf g}^{(6)} = A_{l} \oplus X_{6 - l}\\
                1,              &  {\bf g}^{(6)} = D_{l} \oplus X_{6 - l}\\
                \frac{13}{4},   &  {\bf g}^{(6)} = E_{6} \\
        \end{array}
        \right.
\label{Gposdef}
\eq
These bounds must be compared to the  minimum length that the Wilson
line needs in order to break ${\bf g}^{(16)}$ to a given subalgebra.
For all subalgebras that can be constructed by removing one dot
from the extended Dynkin diagram this length can be calculated very
simply by looking at the simple ideal of the subalgebra to which the
highest root of ${\bf g}^{(16)}$ belongs. The standard Wilson line
is then the dual of this vector with respect to the subalgebra and its
length can be read off the inverse Cartan matrix of the subalgebra.
The highest root can be part of a A-type Lie algebra or of a
D-type Lie algebra. In the first case it is the first or last root
of this Lie algebra and has a norm $<1$: Equation (\ref{Gposdef})
can then be fulfilled for arbitrary ${\bf g}^{(6)}$. In the
second case the highest root of ${\bf g}^{(16)}$ is always the
last or (equivalently) the next to last root of the D--type
subalgebra. Then  ${\bf a}^{2}$  $\geq 1$ if the  the rank of the
summand is $\geq 4$. One can then look for nonstandard solutions
of the symmetry breaking problem. For example the difference
of the last and first to last root is such a solution with norm
squared 1. As the shortest weight vectors of $D_{n}$ Lie algebras
have length 1 shorter solutions cannot exist. We conclude that
this case of symmetry breaking is incompatible with an ${\bf g}^{(6)}$
of the type $D_{l} \oplus X_{6-l}$. This shows that even in the case
of one single Wilson line incompatible patterns of extended
and broken symmetry do exist.

\section{The shift vector realization}
The shift vector method \cite{SW}, \cite{LSW} is, like our construction of
critical Wilson lines a procedure that  starting from one model produces
a discrete set of models and makes it possible to project out
or bring in some states. Therefore it
is natural to ask for the relation between the two methods.
Although the shift vector method was mostly used in covariant
lattice theories we can use it in our context because all Narain
lattices can be embedded in covariant lattices and if we restrict
the shift vector to operate only on the Narain part we can stay
inside the Narain model. The analysis will be done for critical
Wilson lines that only exchange two basis vectors; this includes
all breakings to maximal subalgebras.

The essentials of the shift vector method are the following:
We start with some given even selfdual lorentzian lattice $\Gamma$
$\equiv$ $\Gamma_{p;q}$ and choose some vector ${\bf s}$ $\not\in$
$\Gamma$, such that $N{\bf s} \in \Gamma$, for some minimal positive
integer $N$. ${\bf s}$ defines the ``invariant sublattice"
\be     \Gamma_{0} = \{ {\bf v} \in \Gamma | {\bf v}\cdot {\bf s}
        \in \Zzahl \}.
\eq
If ${\bf s}^{2} = 2m/N$, for some $m \in \Zzahl$, the lattice
\be     \wt{\Gamma} = \sum_{k=0}^{N-1}(\Gamma_{0}+ k{\bf s})
\eq
is even selfdual if and only if ${\Gamma}$ is. Shift vectors can
be applied successively in order to project out unwanted or to
bring in wanted states (more properly: conjugacy classes of lattice
vectors modulo the underlying root lattice). Successive shift vectors
are compatible, that is the ordering is irrelevant, if and only if
their scalar product is an integer. All compatible shift vectors are
then elements of the new lattice.

Comparing this prescription to the relation between the two basis
(\ref{BIGOHF}) and (\ref{newbasis}) connected by the switching on
of background fields it is natural to
assume that the new basis vectors (those basis vectors of $\Gamma'$
that  are not in $\Gamma$)
can be interpreted as successive
shift vectors. Of course all new basis vectors must be {\em rational}
linear combinations of the old ones so that some multiple of each
new basis vector is in the old lattice. This can be achieved by
allowing only rational values for the background fields. Then the
new basis vectors are admissible and compatible shift vectors because
they all have even norm squared and mutual integer scalar product
by construction (see (\ref{SPIG})).
What must be proven is that the lattices $\wt{\Gamma}$, constructed
by the successive application of the new basis vectors as shift vectors
and $\Gamma'$, generated by (\ref{newbasis}) are identical.
All basis vectors of $\Gamma'$ are in $\wt{\Gamma}$ because they
are either shift vectors or invariant vectors of the old lattice
$\Gamma$, implying:
\be   \Gamma' \subset \wt{\Gamma}.
\eq
Dualizing this equation gives
\be   (\Gamma')^{*} \supset  (\wt{\Gamma})^{*}
\eq
But both lattices are are selfdual and so we have the other inclusion
and thus equality.

This result shows how one can keep the information about continuous
parameters when working with shift vectors. If one is only interested
in the gauge group it is preferable to work with a basis instead
of the complete list of conjugacy classes as is usually done when
applying shift vectors \cite{LSW}. The full information can be restored,
 if needed,
by adding the basis vectors modulo the root lattice of the underlying
Lie algebra, until the full group of conjugacy classes of the lattice
\cite{LSW} is generated.

\section{Further applications}
The results and techniques established here do not only enlarge
the knowledge about toroidal compactifications but will also be
helpful in more general situations. One future application is
the construction of toroidal orbifolds with continuous Wilson
lines, which is possible when the twist is realized as a rotation both
in the compactification and in the gauge sector \cite{INQ}, \cite{IMNQ}.
Using the
parametrization of general background fields introduced here
I have been able until now to bring the compatibility condition
between a symmetric twist and general background fields in a simple
form and to calculate the number of untwisted moduli. Combining
the shift vector realization of critical background fields with
the results of \cite{SW} one should be able to reinterpret covariant
lattice models as orbifolds with certain critical background fields.
The usual formalism for orbifolds in which the twist is realized as a
shift \cite{IMNQ} in the gauge sector allows only
for discrete Wilson lines and can be interpreted as a hybrid construction
because it uses both rotations and shift vectors. Some of them
may therefore also be critical points ( or hypersurfaces ) in
the moduli space of the bigger class of orbifolds with continuous
Wilson lines. I hope that the methods developed here will
help to analyze these questions further.

%\bibliography{litbank}

\end{document}